\documentclass[12pt,preprint]{aastex}

\newcommand{\msun}{\rm M$_\odot$}
\newcommand{\re}{\rm R$_{\oplus}$}

\shorttitle{Circumstellar Gas around the DAZ White Dwarf WD 1124-293} 
\shortauthors{Debes et al.}
\begin{document}
\title{Detection of Weak Circumstellar Gas around the DAZ White Dwarf WD 1124-293: Evidence for the Accretion of Multiple Asteroids}
\author{J. H. Debes\altaffilmark{1},M. Kilic\altaffilmark{2}, F. Faedi\altaffilmark{3}, E.~L. Shkolnik\altaffilmark{4},M. Lopez-Morales\altaffilmark{5,6}, A.~J. Weinberger\altaffilmark{7}, C. Slesnick\altaffilmark{7,8}, R.~G. West\altaffilmark{9} }

\altaffiltext{1}{Space Telescope Science Institute, 3700 San Martin Dr., Baltimore, MD 21218}
\altaffiltext{2}{Homer L. Dodge Department of Physics and Astronomy, The University of Oklahoma, 440 W. Brooks St., Norman, OK 73019}
\altaffiltext{3}{Astrophysics Research Centre, School of Mathematics and Physics, Queen's University Belfast, University Road, Belfast, BT7 1NN, UK}
\altaffiltext{4}{Lowell Observatory, Flagstaff, AZ 86001}
\altaffiltext{5}{Institut de Ci\`encies de l'Espai (CSIC-IEEC), Campus UAB, Facultat de Ci\`encies, Torre C5, parell, 2a pl, E-08193 Bellaterra, Barcelona, Spain}
\altaffiltext{6} {Visiting Investigator, Department of Terrestrial Magnetism, Carnegie Institution of Washington, 
5241 Broad Branch Road. N.W., Washington DC 20015, USA}
\altaffiltext{7}{Department of Terrestrial Magnetism, Carnegie Institution of Washington, 5249 Broad Branch RD, N.W., Washington, DC 20015}
\altaffiltext{8}{Charles Stark Draper Laboratory, Inc., 555 Technology Square, Cambridge MA 02139} 
\altaffiltext{9}{Department of Physics and Astronomy, University of Leicester, University Road, Leicester, LE1 7RH, U.K.}
\begin{abstract}
Single white dwarfs with photospheric metal absorption and no dusty disks are believed to be actively accreting metals from a circumstellar disk of gas caused by the disruption and sputtering of asteroids perturbed by post-main sequence planetary systems.
We report, for the first time, the detection of circumstellar Ca~II gas in absorption around the metal-polluted white dwarf WD 1124-293 at $>$7 $R_{\rm WD}$ and $<$32000~AU, with a probable distance of $\sim$54~R$_{\rm WD}$.  This detection is based on several epochs of high-resolution optical spectroscopy around the Ca~II H and K lines ($\lambda$=3968\AA, 3933\AA) with the MIKE spectrograph on the Magellan/Clay Telescope.  We confirm the circumstellar nature of the gas by observing stars with small angular separations to WD~1124-293 and larger distances from Earth--none of the reference stars show absorption consistent with the presence of appreciable local interstellar medium at the distance of the white dwarf.  By combining our observations over four years with previous spectra of WD~1124-293, we have measured the equivalent width of the two photospheric Ca lines over a period of 11 years.  We see $<$ 5-7\% epoch-to-epoch variation in equivalent widths over this time period, and no evidence for long term trends in the strength of the Ca~II lines.  Since it is likely that the circumstellar gas arises from a disk edge-on to our line of sight, we also place limits to short period transiting substellar and planetary companions with R $>$ R$_{\rm \oplus}$ using the WASP survey.  The presence of gas in orbit around WD~1124-293 implies that most metal rich WDs could harbor relic planetary systems.  Since roughly 25-30\% of WDs show metal line absorption, the dynamical process for delivering small bodies close to a WD must be robust.

\end{abstract}


\section{Introduction}
In the absence of externally accreted material, radiative levitation, or convective dredge-up, white dwarfs (WDs) should have pure hydrogen or pure helium photospheres.  All other elements heavier than helium should settle out of view on timescales much shorter than the cooling age of WDs.  Hydrogen WDs (type DA) have photospheric settling times that can range from days to thousands of years, while helium WDs (type DB) have settling times as long as 1 Myr.  Due to their high gravity and thin photospheres, WDs are some of the most sensitive probes to low levels of external accretion.  A typical 0.6\msun\ hydrogen WD with an effective temperature of 10000~K can show observable Ca~II lines from the accretion of $\sim$10$^{6}$~g/s of solar abundance material, corresponding to  the accretion of a small asteroid's worth of material per year \citep{paquette86,koester05,koester09}.  WDs are sensitive probes of the accretion from the
 surrounding interstellar medium (ISM) \citep{dupuis92,dupuis93,dupuis93b}, the winds from companions \citep{sion84,zuckerman03,debes06}, and the accretion of asteroidal material from orbiting dust disks \citep[e.g.][]{zuckerman07,klein10,vennes10,dufour10,debes11a,debes11b}.  All three of these phenomena can explain the presence of metal lines in WDs, but a significant fraction of metal polluted WDs (type DAZ for hydrogen atmosphere WDs and type DBZ or DZ for helium atmosphere WDs) are apparently single, do not possess any dust disks, and do not have sufficient surrounding ISM to explain their accretion \citep{aanestad93,kilic07}.

ISM accretion was the first logical explanation for metal polluted WDs \citep{wesemael78,wesemael82,dupuis92,dupuis93,dupuis93b}.  However, studies of polluted WDs show no strong correlation between velocity and metal abundance as would be expected for Bondi-Hoyle accretion of the ISM \citep{aanestad93,zuckerman03,zuckerman10,kilic07}.  Nor is there any correlation between metal abundance and constraints on the local ISM \citep{kilic07}.  Most observed DAZs and DZs reside well within the low density Local Bubble, and are not physically located near known dense molecular clouds.  The lack of appreciable hydrogen in DZs also cannot easily be explained by the accretion of solar abundance material.

The presence of polluted DAZs that are members of WD+red dwarf binaries raises the possibility that previously unseen companions could be the cause of many polluted WDs through the accretion of a companion wind \citep{zuckerman03}.  While it is physically plausible that a close ($a <$1~AU) companion can pollute a WD with its wind \citep{debes06}, most WDs that have been observed with {\em Spitzer} are constrained to have no close stellar or brown dwarf companions  \citep{mulally08,debes07,farihi09}.  The most massive objects that could orbit these WDs would have masses of 5-15~M$_{\rm Jup}$.

Dusty WDs all show pollution due to the direct accretion of metal rich grains onto the WD surface.  However, only $\sim$30\% of metal polluted WDs show infrared excesses due to dusty disks \citep{kilic08,farihi10}.  Some disks may be close to edge-on to our line of sight and hidden, but inclination effects alone cannot explain the relative number of dusty and non-dusty WDs.  Tidally disrupted planetesimals perturbed by planets are believed to be the origin of dusty WDs \citep{debes02,jura03,bonsor11}.   
Similarly, apparently single WDs with metal pollution may be caused by hitherto unseen disks that are either dusty, optically thin, and undetectable in the infrared, or entirely gaseous.  Gaseous disks with some mass in gas of $M_{\rm disk}$ that arise from the sputtering of debris from tidally disrupted planetesimals have been postulated to explain the relative number of dusty and non-dusty WDs \citep{jura08}.  In this scenario, only planetesimals larger than $\sim$2~M$_{\rm disk}$ create an optically thick dusty disk, while smaller objects are completely ablated by the relative velocity between the debris and the existing gaseous disk, further populating the disk with gaseous material.

Circumstellar gas emission lines have been observed toward several hotter WDs that also possess dust disks \citep{gaensicke06,gaensicke07,gaensicke08,melis10}, demonstrating that significant gas can be generated from either the sublimation, collision, or photoionization of dust grains.  Similarly, a handful of hotter WDs that have been observed in the UV show circumstellar gas absorption \citep{holberg98,bannister03, lallement11,dickinson12}.  These hotter WDs may be evaporating circumstellar dust disks or local ISM, but to date none of these systems has been shown to possess a dust disk--their origins are less clear.  As WDs become hotter, radiative levitation of metals becomes a more important process and can explain the presence of metals in a WD atmosphere for effective temperatures larger than $\sim$25000~K \citep{chayer10}.

In this work we present the detection at 8-$\sigma$\ of weak circumstellar Ca~II absorption in orbit around the cooler DAZ WD~1124-293.  Based on observations of other sight lines close to WD 1124-293 and the relative velocity of the circumstellar gas, we confine the spatial extent of the absorption to 7~$R_{\rm WD} < r <$ 32000~AU from WD~1124-293.  We argue that it must arise from a circumstellar gaseous disk in orbit around the WD at a probable radius of $\sim$54~$R_{\rm WD}$, well within the WD's tidal disruption radius.  In \S \ref{sec:obs} we describe the MIKE spectroscopic observations of WD~1124-293 and present the results of those observations in \S \ref{sec:res}.    We also search WASP data for planetary transits around WD~1124-293 in \S \ref{s:trans}.  We search for variability in the accretion of material on the WD in \S \ref{sec:var} and present our conclusions in \S \ref{sec:conc}.

\section{Observations}
\label{sec:obs}

\begin{figure}
\plotone{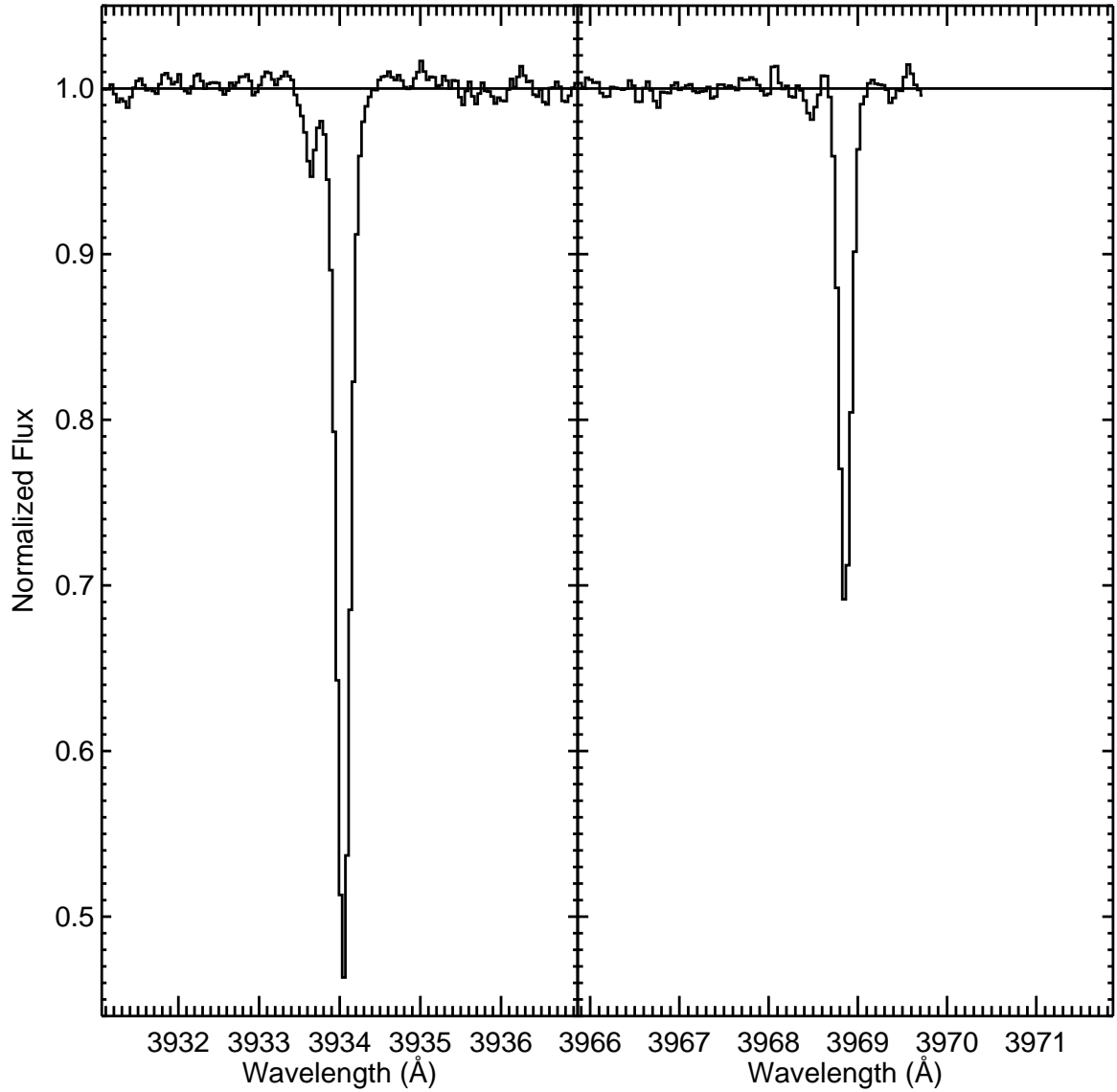}
\caption{\label{fig:f1}Detection of weak circumstellar Ca~II features in the spectrum of WD~1124-293.  The spectrum is from a S/N weighted combination of observations spanning four years, and has a S/N of 180.  The photospheric line is located at $\approx$3934\AA\ and the circumstellar feature is blue shifted from the photospheric line.}
\end{figure}

\begin{figure}
\plotone{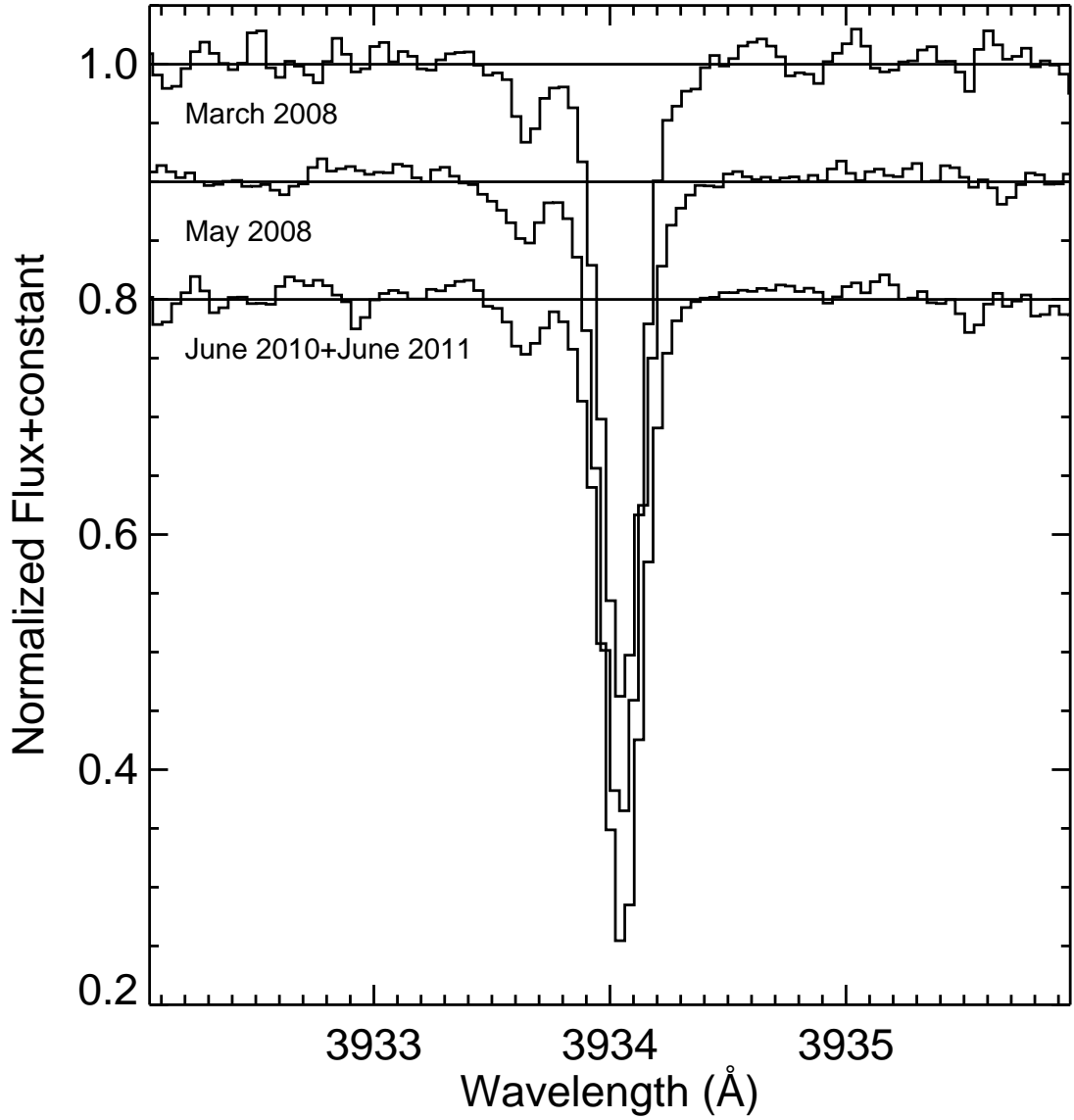}
\caption{\label{fig:f2} Three separate epochs which show a clear detection of the blueshifted weak Ca~II feature.  The line is recovered relative to the continuum with a significance of 4.7-, 4.9-, and 4-$\sigma$ for March 2008, May 2008, and June 2010+June 2011 respectively.}
\end{figure}

\begin{figure}
\plotone{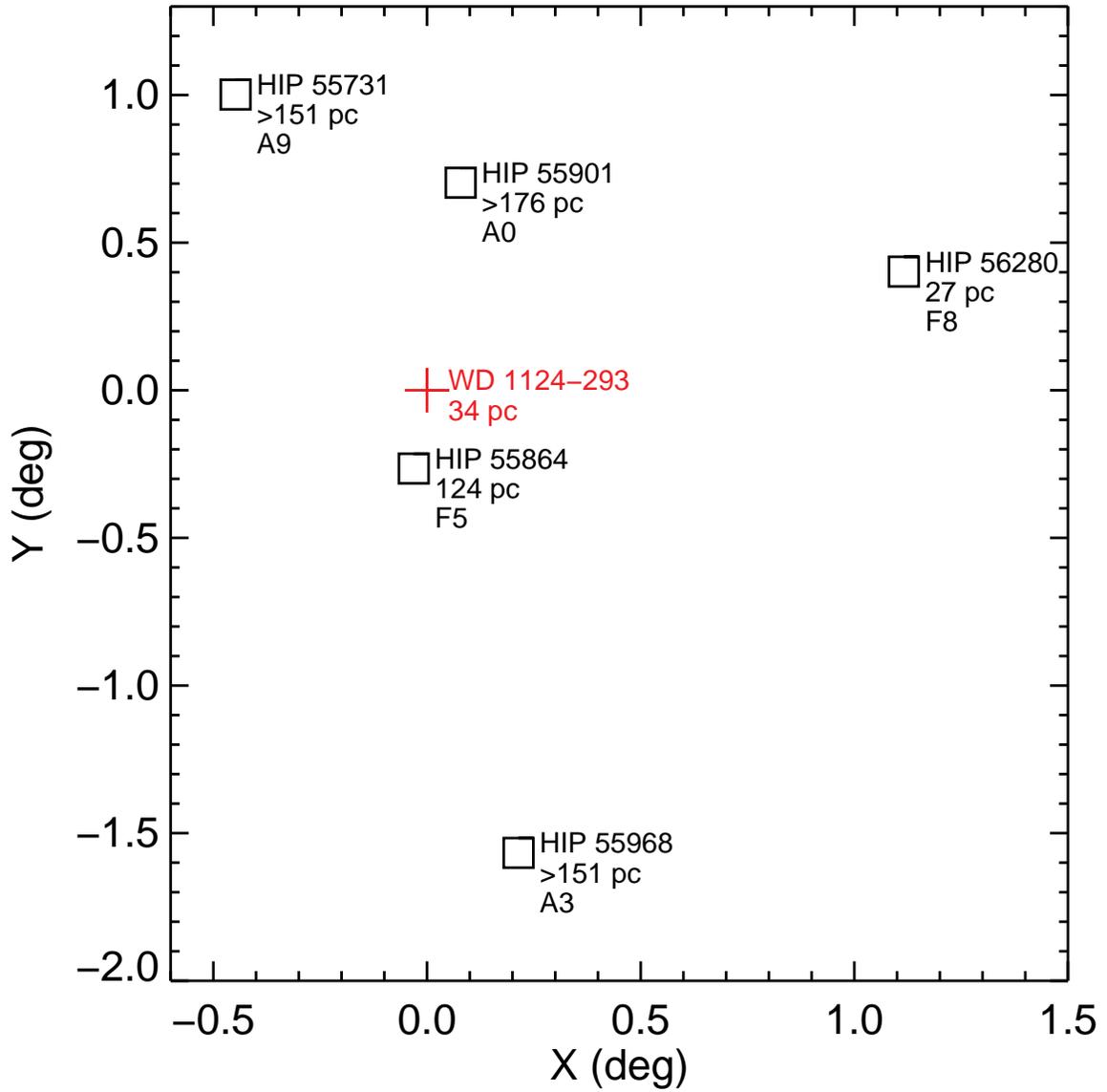}
\caption{\label{fig:f3} Location of ISM standards on the sky relative to WD~1124-293.  HIP~55864 is the closest star to WD~1124-293 and is at a projected separation of 0.16~pc at the distance to the WD.}
\end{figure}

We have taken optical spectra of WD~1124-293 over 16 separate days spanning 4 years with the MIKE spectrograph \citep{bernstein03} on the Clay Telescope at Las Campanas Observatory between 2007 and 2011 (See Table \ref{tab:obs}) as part of a large DAZ monitoring project \citep{debmuk10}.  WD 1124-293 was first discovered in the Edinburgh-Cape Survey and listed in \citet{mccook99}.  Based on spectral models of WDs, WD~1124-293 has M=0.66 \msun\ and T$_{\rm eff}$=9400~K \citep{koester06}, with a spectroscopically determined distance of 33.6~pc \citep{pauli06}. Photospheric metal lines of Ca II were detected with moderate resolution spectroscopy from Keck HIRES (R$\sim$30000) and VLT UVES (R$\sim$18500) \citep{zuckerman03,koester05}, which showed moderately strong Ca II H and K lines ($\lambda$=3968.469,3933.664) with a heliocentrically corrected redshift of 29.5~km~s$^{-1}$ (29$\pm$1~km~s$^{-1}$) for Keck (VLT) in December 1998 and April 1999 (April and May 2000).  The calcium cannot be explained by accretion of the local ISM based on measurements of the local ISM density \citep{kilic07}.  The equivalent width (EW) of the Ca~II line over the four separate epochs with VLT and Keck averaged $\approx$106~m\AA\ with no significant variation over this period, implying a steady state accretion rate of 4$\times10^{15}$ g/yr {\citep{koester06}, assuming 1\% metal rich material relative to solar abundance.  In January and June of 2010 we observed four bright F and A stars located near WD~1124-293 on the sky in order to probe the surrounding ISM at the distance of WD~1124-293.

For our MIKE spectra, we used a 0\farcs7$\times$5\arcsec\ slit which corresponded to a resolution of $\sim$40,000 at the Ca II K line.  Th-Ar comparison lamp spectra were taken near in time to each spectrum of WD~1124-293.  In addition, we obtained the previously reduced spectra of this WD from the VLT and Keck \citep{koester05,zuckerman03}.  The VLT spectra were kindly provided by D. Koester, while the Keck data was publicly available with the journal article it was published in.

Our MIKE data were flatfielded and bias-subtracted using the MIKE reduction pipeline written by D. Kelson, with methodology
described in \citet{kelson00} and \citet{kelson03}.  We initially extracted the spectra using the full pipeline, but discovered that the pipeline was subtracting the background at a level of precision insufficient for high quality equivalent width measurements.  To rectify this, we performed our own extraction, background subtraction, and wavelength solution for 8-10\AA\ windows centered on the Ca H and K lines for all of our MIKE spectra.  The spectra were also corrected for heliocentric velocity shifts between observations.  Since the Ca II K line falls on two orders, we combined each order by fitting the continuum with a polynomial and averaging the orders for each night to increase signal-to-noise (S/N) to the values reported in Table \ref{tab:obs}.  All the fitted spectra from MIKE were then averaged with a S/N weighting to produce a final spectrum of the region around each line.

In order to check for variability in our spectra between 2007-2011 as well as with the previous observations, we calculated the EW of the two calcium lines from each night's spectrum. To calculate the EW, we chose a window around the line equivalent to $\pm$3 times the standard deviation of the line as determined by a Gaussian fit, which also returned a measure of the radial velocity for the WD on each night.  We chose several different polynomial fits to the continuum and added any systematic uncertainty in the continuum fit to our uncertainties in the EW calculation.  This approach has the advantage of calculating a more accurate EW despite differences in resolution between different epochs, since the total integrated line flux is conserved from observation to observation.  Table \ref{tab:meas} details the observed photospheric line radial velocities and equivalent widths.  From the individual measures of the K line with MIKE, we obtain a median radial velocity for WD~1124-293 of 29.0$\pm$0.5~km~s$^{-1}$.

\section{Results}
\label{sec:res}

The final, S/N weighted spectrum of WD~1124-293 at the Ca H and K lines is shown in Figure \ref{fig:f1}.  Just to the blue of the main photospheric lines we detect weak lines indicative of another Ca~II component, with line depths of 3.4- and 8.5-$\sigma$ below the continuum for the H and K line respectively.  As seen in Table \ref{tab:meas}, the lines at both wavelengths are separated by $\approx$30~km~s$^{-1}$ to the blue of the photospheric Ca~II.  Weak lines in the spectrum could be due to additional photospheric species, circumstellar Ca~II gas, intervening ISM Ca~II gas, or an artifact from the spectrograph or reduction process.  In this section we argue that the lines arise from circumstellar Ca~II gas in orbit around the WD.

\subsection{Ruling out artifacts and other atomic species}

There are three separate conditions that rule out the possibility that the lines could arise from the data reduction process or the MIKE spectrograph.  First, we have verified that the line near the photospheric Ca~II K line was detected significantly on a S/N combined spectrum of each of the two echelle orders that the line fell on.  The position of the line on the two orders is separated by $\approx$1100 pixels, ruling out possible artifacts from the instrument.  For both orders, the line is detected at the proper position and at 4- and 6-$\sigma$\ below the continuum.  Secondly, the line is detected significantly in the final combined spectra when using the full Kelson MIKE pipeline, which used different methodology to perform order extraction and background subtraction.  Finally, when we combine the data into subsets, we recover the line as well.  Figure \ref{fig:f2} shows three separate epochs: March 2008, May 2008, and a combination of our June 2010 and June 2011 data.  In particular, the heliocentric velocity correction in March 2008 and our June 2010+2011 data differed by 23 km~s$^{-1}$, ruling out any terrestrial origin for the line, since the epoch-to-epoch centers match to within 0.01\AA.

These lines could be weak photospheric features.  We searched atomic line lists, as well as metal lines other than calcium reported for other polluted WDs \citep[e.g.][]{klein10}, and found no atomic lines that could plausibly be matched to the observed spectrum of WD~1124-293.  Furthermore, the line depths and relative equivalent widths are consistent, within the uncertainties, to the expected line strength ratio between the Ca~II H and K lines.  Based on these arguments, the weak features must arise from intervening ISM or circumstellar Ca~II gas.

\subsection{Ruling out Local Interstellar Absorption}

While ISM absorption could explain the presence of the weak features we observe, we rule out an ISM contribution.
 The Sun resides in a low-density ISM structure
known as the Local Bubble. WD 1124-293 is at a spectroscopically determined distance of 33.6 pc, well inside the Local Bubble \citep{pauli06}.
The local interstellar medium is relatively devoid of cold neutral gas, until
the accumulation of dense material at $\sim$80 pc \citep{welsh10}. However, warm, partially ionized clouds
have been detected through weak \ion{Ca}{2} ISM absorption features in the spectra of stars within 50 pc
of the Sun \citep[see][for a review of the properties of the local interstellar medium]{redfield08,welsh10}.

If the weak Ca~II absorption feature in WD 1124-293 is due to ISM accretion, other stars in close proximity on the sky
should also show a similar absorption feature. In order to test this possibility,
we observed several bright stars at angular separations $<1.6\arcdeg$ from WD 1124-293 to search
for absorption features at the same velocity. This technique, as well as searching for line
variability, has been used in the past to confirm circumstellar absorption around HD 32297,
a young star with an edge-on debris disk \citep{redfield07}. We observed four stars with constraints on their distance from Hipparcos and with the same setup
as WD 1124-293: HIP 55864 (F5V, r=0.27$\arcdeg$) HIP 55901 (A0V, r=0.71$\arcdeg$), HIP 55731 (A9V, r=1.10$\arcdeg$),
 and HIP 55968 (A3V, r=1.58$\arcdeg$) (See Table \ref{tab:obs}}).
Another star, HIP 56280A (F8V, r=1.19$\arcdeg$), had a spectrum that was kindly provided
to us by S. Desidera.  Figure \ref{fig:f3} shows how each ISM standard is arranged on the sky relative to WD~1124-293.

The distances for each star are given in the latest reduction of Hipparcos data \citep{leeuwan07}.  HIP~56280A and HIP~55864 have secure distances (parallax measurements with errors of $<$20\%), while HIP~55731 and HIP~55901 have marginally detected parallaxes ($\sim$3-$\sigma$).  HIP~55968's parallax is consistent with zero.  For the marginal detections, we take as the stellar distance the 3-$\sigma$ upper limit to the parallax, which is 6.62~mas (151~pc) and 5.67~mas (176~pc) for HIP~55731 and HIP~55901 respectively.  HIP~55968's optical and NIR magnitudes and spectral type are consistent with this star lying between HIP~55731 and HIP~55901, so we take the lower limit to its distance to be 151~pc.

\begin{figure}
\plotone{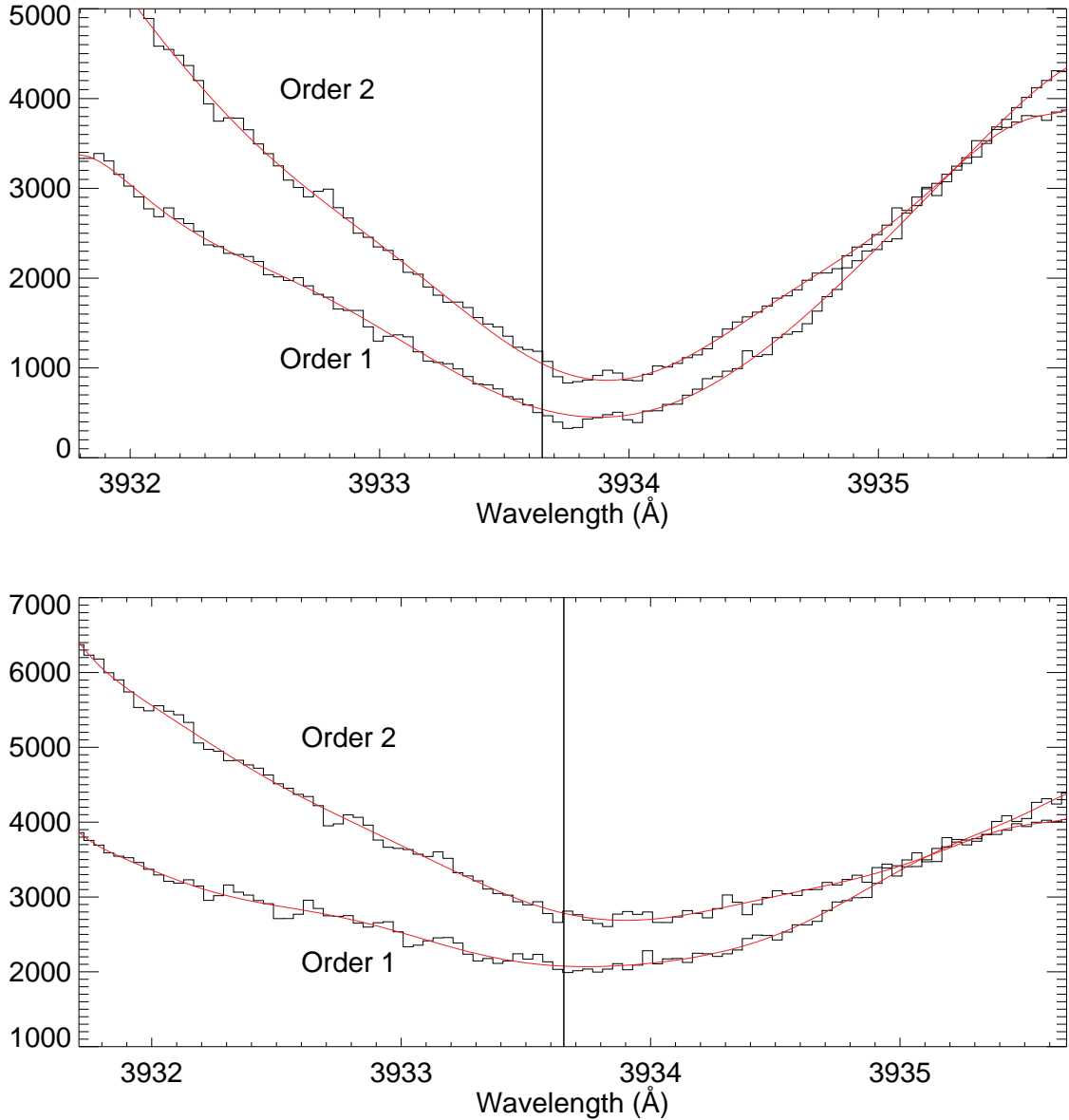}
\caption{\label{fig:f4} Comparison of polynomial fitting of the broad and deep photospheric Ca~II lines of two ISM standards, HIP~55731 and HIP~55864.  The histograms are the observed spectra in each order of the MIKE echelle spectrum while the solid lines are the fits.  The solid vertical line marks the location of WD~1124-293's circumstellar feature.}
\end{figure}

\begin{figure}
\plotone{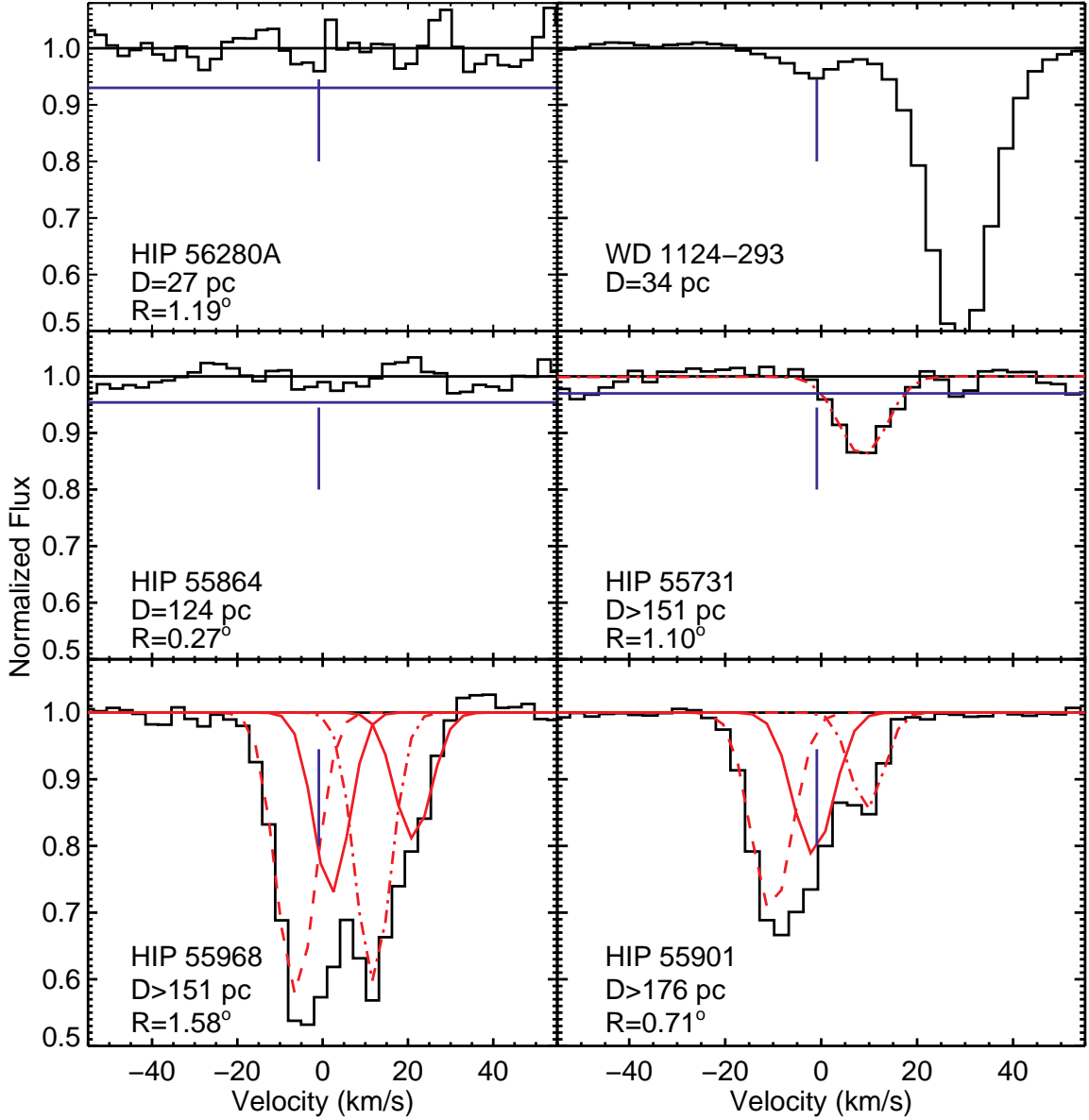}
\caption{\label{fig:f5}Comparison of the location of the weak Ca~II K feature in WD~1124-293 with our ISM standards, arranged by probable distance from Earth.  The dark vertical line shows the location of the weak circumstellar feature of WD~1124-293, while the horizontal blue line shows the 3-$\sigma$ detection threshold for standards that showed no absorption at the wavelength of WD~1124-293's component.  More distant standards show ISM absorption, while standards closer than 124~pc show no appreciable ISM absorption, ruling out LISM as a source for WD~1124-293's blueshifted feature.}
\end{figure}

Each star had broad Ca~II K photospheric absorption which we removed by applying a polynomial
fit to the photospheric line, an example of which is shown in Figure \ref{fig:f4}.  For HIP~56280A, its narrow line core was fitted with two polynomials joined at the line core center of 3933.704\AA.  The resulting continuum fitted spectra were inspected for narrow absorption components.  HIP~56280A (to lower significance) and HIP~55864 show no appreciable absorption features comparable in strength to the WD~1124-293 line, while HIP~55731 shows a weak (presumably ISM) component far from the radial velocity or our observed feature.  HIP~55901 and HIP~55968, both constrained by Hipparcos to be more distant than HIP~55864, each showed absorption from at least three ISM components.  We attempted to simultaneously fit these three components for the two stars and present our best fits in Figure \ref{fig:f5}, which shows WD~1124-293's Ca~II K line region compared with our ISM standards.  Table \ref{tab:ism} shows the resulting velocity components for these fits, consistent with at least three distinct complexes of ISM gas that slowly vary on size scales of $\sim$1-2~pc, one of which is at a radial velocity within a few km~s$^{-1}$ of the circumstellar feature.  Despite the similar radial velocity of the two features to what is observed in WD~1124-293, the non-detection of any similar lines for HIP~56280, HIP~55864, and HIP~55731 constrains the location of these ISM components to be at a distance $>$151~pc.  The non-detection of the faint candidate circumstellar component for HIP~55864
constrains the extent of the Ca~II gas near WD 1124-293 to $<0.27\arcdeg$ at d=33.6 pc, or roughly
32000 AU (0.16 pc).

\citet{welsh10} find a slowly increasing \ion{Ca}{2} equivalent width with increasing
distance, but depending on whether local gas clouds are encountered or not, they find
both low and high values of \ion{Ca}{2} volume density within 30 pc. Out of the 1857 stars
analyzed by \citet{welsh10}, 50 are within 40 pc of the Sun. Only one of these 50 stars,
HD~159561 ($l=36\arcdeg, b=+23\arcdeg$), shows an ISM absorption feature with a \ion{Ca}{2}
equivalent width stronger than what is observed for the weak feature in WD1124-293
($l=282\arcdeg, b=+30\arcdeg$). The ISM complexes observed for our more distant stars are consistent with a relatively dense cloud of material that lies $>$90~pc along the Galactic sightline to these stars \citep{welsh10}, and thus cannot account for the weak Ca~II feature we measure for WD~1124-293.  Hence, it is unlikely that this feature is due to an intervening warm ISM cloudlet.  The only scenario that the data allows is if an extremely narrow warm ISM column near WD~1124-293 (width $\sim$0.06~pc and height $\sim$1.3~pc) with a velocity gradient intersected the sightlines to WD~1124-293, HIP~55901, and HIP~55968, but did not intersect the sightline of HIP~55864.  

Such a pathological structure is not supported by studies of the nearby ISM.   \citet{redfield01} studied the small scale (0.05-1.2 pc) structure of the local
interstellar cloud and find that the \ion{Mg}{2} column densities do not vary by more than a factor of
two for $\le0.6$~pc scales.   The implied Ca~II column densities by the much stronger ISM absorption observed towards HIP~55901 and HIP~55968 are an order of magnitude larger than for WD~1124-293, and would have been easily detected in the ISM standards that showed no component at the velocity of our candidate circumstellar feature.  Hence, the observations of our ISM standards limits the amount of interstellar Ca~II between WD 1124-293 and us at the velocity of the Ca~II K feature to EW $<$ 6~m\AA.

The source of metals in high surface gravity WDs has been a puzzle for decades \citep[see][]{dupuis93}.
The search for a correlation between the local ISM clouds and the locations of metal-rich WDs was
inconclusive \citep{zuckerman03}.  \citet{kilic07} used the observed ISM column densities toward stars
in close proximity to the known DAZ WDs to demonstrate that there is no correlation between the accretion
density required to supply metals observed in DAZs with the densities observed in their interstellar environment, including WD~1124-293.
Therefore, the DAZ WD population as a whole argues against ISM accretion being
the dominant contributor to the metals in WDs.

In conclusion, it is more likely that the weak Ca~II feature is due to circumstellar gas in orbit
around WD1124-293 because (1) no similar absorption features are detected in
neighboring sightlines as close as 0.16~pc and the local interstellar medium is relatively homogeneous
at such scales \citep{redfield01}, (2) the strength of the absorption feature is uncommon for
stars within 40-50 pc of the Sun, (3) the absorption feature matches the stellar radial velocity,
and (4) there is no correlation between the metal abundances of known DAZ stars and the ISM densities
\citep{kilic07}.

\subsection{Limits to the inner extent of circumstellar gas}
If the line arises from gas in orbit around the WD, the relative velocity between the photospheric line and the circumstellar feature can be used to determine how deep the gas lies in the gravitational well of the WD.  The photospheric Ca~II line is offset from the true systemic velocity of the WD due to the gravitational redshift of WD~1124-293.  Using the latest values of WD~1124-293's $\log$~g=8.096 and its T$_{\rm eff}$=9420$\pm$150~K (D. Koester, private communication), we have calculated masses from cooling models of WDs\footnote[1]{These models are available at http://www.astro.umontreal.ca/$\sim$bergeron/CoolingModels} \citep{bergeron95,holberg06,tremblay11,bergeron11} which use the mass-radius relation of \citet{fontaine01} to determine the gravitational redshift of WD~1124-293.  The uncertainty in the gravity of the WD dominates the uncertainty in the expected gravitational redshift, but is also not well known.  To estimate this possible uncertainty we compare our value to that given in \citet{koester01} and \citet{koester09}, which vary between 8.04 to 8.099.  We therefore estimate the systematic uncertainty in the WD gravity to be $\approx$0.06.  Using this value and uncertainty and calculating a gravitational redshift ($v_{\rm grav})$, we obtain  $v_{\rm grav}$=34.9$\pm3.7$~km~s$^{-1}$, implying a systemic velocity $\gamma$=-5.9$\pm3.7$~km~s$^{-1}$

Previous calculations of WD~1124-293's $\gamma$ (based on its inferred mass and radius from synthetic spectral modeling) find -2.7$\pm$4.3~km~s$^{-1}$, consistent with our calculated $\gamma$ \citep{pauli03,pauli06}.  The primary difference between our result and previous calculations of $\gamma$ come from a higher WD gravity (The earlier values of $\log~g$ being 0.05 dex lower), highlighting the dependence of this calculation on a detailed knowledge of a WD's properties.  The measured relative velocity between the photospheric line and the circumstellar line ($v_{\rm gas}$) from our S/N weighted spectrum is -29.9$\pm$0.8~km~s$^{-1}$.  The orbital radius of the gas is then $R_g$=$v_{\rm grav}/(v_{\rm gas}+v_{\rm grav})$ R$_{\rm WD}$, implying a minimum radius to the gas of 7$^{+11}_{-3}~R_{\rm WD}$.  Given the uncertainties, the faint line could correspond to the systemic velocity and reside further from the WD, thus directly probing the gravitational redshift of the WD.  This minimum radius then should be treated as a lower limit to the gas radius.  For optically thin silicate dust, the sublimation radius is given by \citep{jura08,kimura02,debesbook11}:

\begin{equation}
R_{\rm sub}=3.7\times10^{10} \left(\frac{L_{\rm WD}}{0.001~L_\odot}\right)^{1/2}\left(\frac{1000~K}{T_{\rm sub}}\right)^2~\rm cm.
\end{equation}

From the cooling models used to determine WD~1124-293's gravitational redshift, we also determine a luminosity of almost exactly 0.001~$L_\odot$, leading to a sublimation radius of $\approx$44~$R_{\rm WD}$, which is close to the inferred radius of the gas from kinematic arguments in \S \ref{sec:disc}.

\subsection{Properties and Origin of the Circumstellar Gas}
\label{sec:disc}
Further characterization of the Ca~II lines permits additional constraints on the origin and properties of the circumstellar gas around WD~1124-293.  In particular, the different constraints point to a localized disk of gas a few tens of WD radii from WD~1124-293 and well within its tidal disruption radius.

\begin{figure}
\plotone{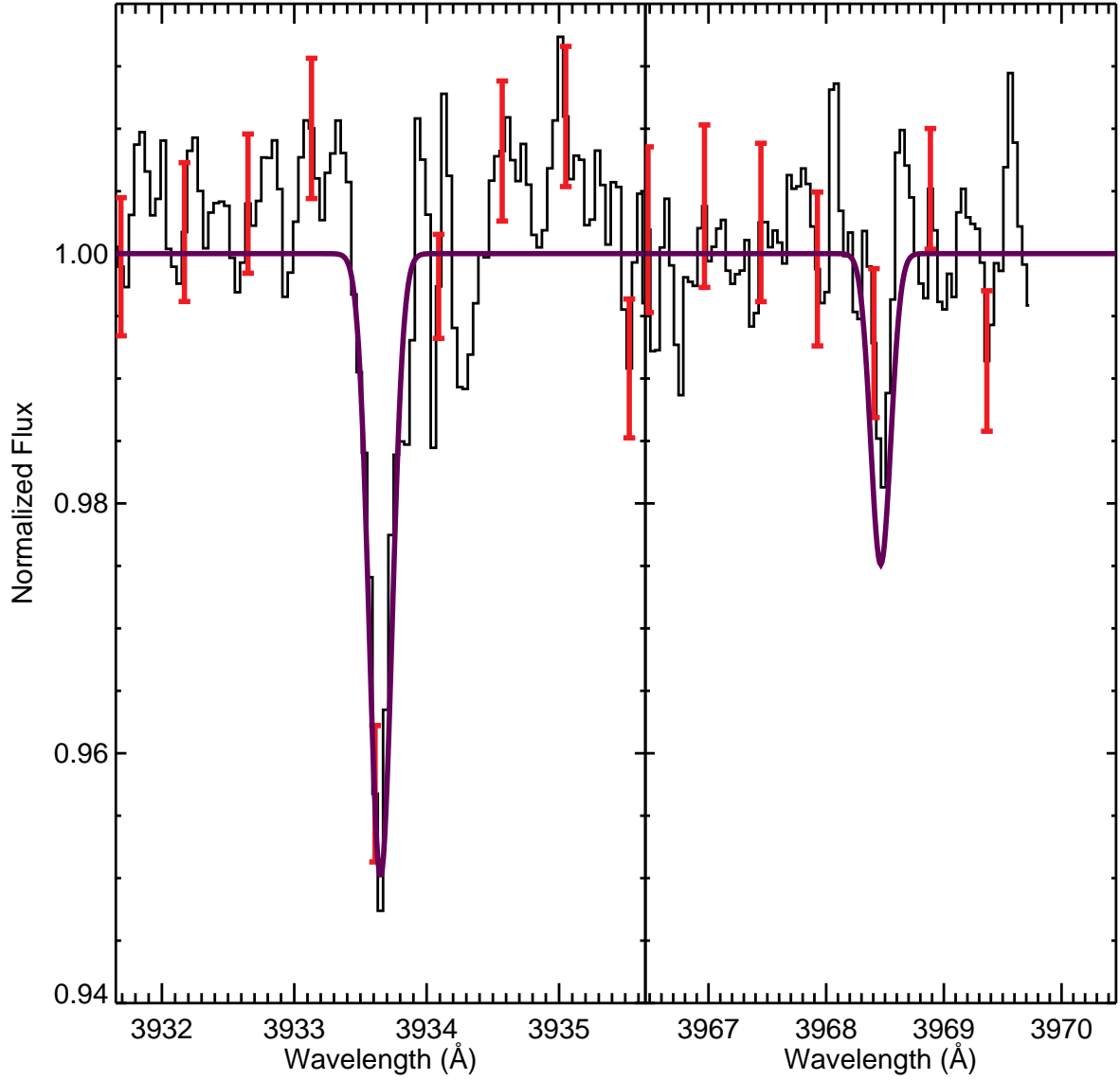}
\caption{\label{fig:f6}Comparison of the observed circumstellar Ca line (black histogram with red error bars) with the photospheric line divided out to a model Ca II cloud (solid dark line) with a column density of $\log N$=11.07$^{+0.1}_{-0.2}$.}
\end{figure}

In order to determine the column density of the Ca~II gas, we minimized a $\chi^2$ metric for an optically thin gas column by varying the column density ($\log~N$) and FWHM of the line, using the proper oscillator strengths and assuming a Gaussian line profile.  From the minimization we find a best fitting line model with (98\% confidence) $\log N$=11$^{+0.1}_{-0.2}~cm^{-2}$ and FWHM=0.2$\pm$0.1 \AA\ with a $\chi_\nu^2$=1.08.  The best fit line, compared to the observed circumstellar line is shown in Figure \ref{fig:f6}.  The model overpredicts the expected line strength of the Ca~II H line, but given the low S/N of the line it is hard to determine if this difference is significant.  Since this is the ionized state of calcium, it is impossible to tell how much total gas resides in the system without a) other gas species or b) a measurement of Ca~I to determine the temperature and electron density of the gas.  Further characterization of the gas will be the focus of future work, but is beyond the scope of this paper.  

We can also constrain the location of the gas by the width of the weak Ca~II line--an upper limit to the orbital velocity of the gas can be determined from the line FWHM, and thus a characteristic radius for the disk.  We assume that the gas is in a circular orbit and that the line FWHM corresponds to the range of orbital radial velocities within the circumstellar gaseous disk that intersects the angular disk of the WD on the sky:

\begin{equation}
r_{\rm disk} \approx \left(GM_{\rm WD} \right)^{1/3}\left[\frac{2 R_{\rm WD}}{FWHM(\rm cm~s^{-1})}\right]^{2/3}
\end{equation}

From our gravitational redshift calculations, $R_{\rm WD}\approx$8.4$\times 10^8$~cm, and the FWHM of the line is 16~km~s$^{-1}$, giving a radius for the disk of 4.5$\times10^{10}$~cm or $\approx$54~$R_{\rm WD}$.  This is consistent with the gas being close to the WD and within the tidal disruption radius of WD~1124-293, as well as being from the sublimation of optically thin dust.

The constraints on the location of the gas and the presence of Ca~II combined with the lack of any appreciable ISM all point to a similar origin for WD~1124-293 as that posited for WDs with near-IR excesses due to dusty disks.  These observations fit in with the scenario where dusty excesses are caused by massive asteroids or comets that are dynamically perturbed by a planetary system \citep{debes02,bonsor11,bonsor12,debes12} and then tidally disrupted \citep{jura03}.  However, a size distribution of planetesimals will be disrupted over time, and for smaller bodies, sputtering renders most tidally disrupted dust into a gaseous phase quite quickly, forming circumstellar gas disks instead of circumstellar dust disks \citep{jura08,farihi12}.  Circumstellar gas absorption roughly at the same radius as that observed for dusty disks around WDs is also suggestive of a tidally disrupted asteroidal origin for our observations of WD~1124-293.

\section{Limits to Transiting Companions to WD~1124-293}
\label{s:trans}
Given the possible edge-on inclination of this system due to the presence of circumstellar gas, the probability of any detectable planetary companion transiting WD~1124-293 might be higher than a randomly inclined system.  To that end we have gathered all data of WD~1124-293 from the Wide Angle Search for Planets (WASP) transit database.
From previous observations, WD~1124-293 does not possess close-in brown dwarf or stellar companions.  In their comprehensive {\em Spitzer} search for IR excesses due to companions around a sample of WDs, \citet{farihi08} placed a definitive upper limit of 12~M$_{Jup}$ to the presence of any unresolved companions in orbit around WD~1124-293.  

The WASP North and South telescopes are located in La Palma (ING - Canary Islands) and Sutherland (SAAO - South Africa), respectively. Each telescope consists of 8 Canon 200 mm $f/1.8$ focal lenses coupled to e2v 2048$\times$2048 pixel CCDs, yielding a field-of-view of 7.8$\times$7.8 square degrees with a pixel scale of 13.7\arcsec~\citep{pollacco06}.
The WASP observing strategy is such that each field is observed with a typical cadence of the order of 8--10 minutes and typical exposures of 1 minute (30 sec exposure plus over heads). WASP provides good quality photometry with a precision exceeding 1\% per observation in the approximate magnitude range $9 \le {\rm V} \le 12$.

\begin{figure} 

  \plotone{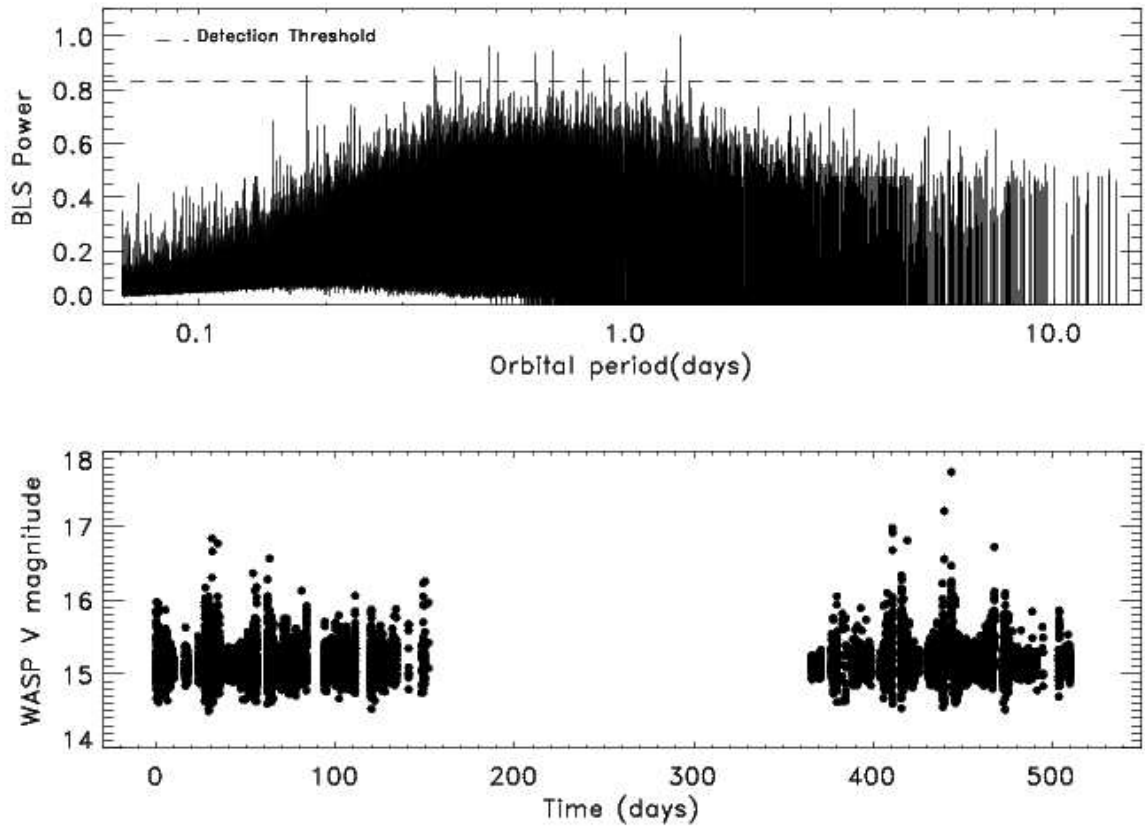}
  \caption{\label{fig:f7}Top panel: Box-Least Square power spectrum for the white dwarf WD1124-293.
    Lower panel: WASP light-curve for WD1124-293 covering the period
    January 2007 - May 2008, with a total of 10618 photometric
    measurments. We used WASP V-magnitude, where
    V$_{mag}=-2.5*log(flux/10^6)$, expressed in $\mu$Vega.}
\end{figure}

WD~1124-293 was observed by the WASP-South telescope
in the 2007 and 2008 seasons covering the interval 2007 January 04 to
2007 June 05, and 2008 January 05 to 2008 May 28, respectively. The
10618 pipeline-processed photometric measurements were de-trended
using the Tamuz algorithm \citep{Tamuz05} to account for
linearly-correlated systematic noise in the data. We show the WASP
light-curve in Figure \ref{fig:f7} (bottom-panel).

We used the modified implementation of the Box-Least Square (BLS) algorithm
described in \citet{Faedi2011} to search for transits and eclipses of
sub-stellar and planetary companions in close orbits around the white
dwarf WD1124-293. We searched a parameter space defined by orbital
periods ranging from 2 hours to 15 days, and companion sizes ranging
from Moon-size to Jupiter-size. No transiting planet has been
identified. Figure \ref{fig:f6} (top-panel), shows the BLS power-spectrum
for the star WD1124-293. The dashed-line indicates the detection
threshold as defined in \citet{Faedi2011} for 10\% noise levels. We have
phase-folded the WASP light-curve to investigate the possible presence
of an astrophysical signal at the periods detected above the
threshold. However, none showed real variation.  
Finally, we have used the modified Lomb-Scargle periodogram
\citep{Scargle82} to search for generic variability. We find that a real power
is assigned to a peak at P=27.42 days which is most probably related
to the Moon cycle.  We used these null results together with
the results of our simulations to estimate an upper-limit to the
frequency of close companions to WD~1124-293.

To assess the chances of detecting eclipses and transits of
substellar and planetary companions to WD1124-293, we
performed an extensive set of Monte Carlo simulations. The approach we
adopted was to create realistic synthetic light-curves containing
eclipse and transit signatures of the expected depth ($\delta_{\rm tr}$) and duration for
a range of companion sizes and orbital periods, then to attempt to
detect these signatures using the modified BLS mentioned above. By
noting the rate at which the BLS search recovered the transit ($f_{\rm det}$) at the
correct period (or an integer multiple or fraction) we were able to
estimate the detection limits of such systems in an automated
manner.

Details of the synthetic transit signal generated, the
corresponding transit probability, depth and duration are described in
detail in \citet{Faedi2011}. Here we present the results of simulations covering
 the orbital period-planet radius parameter
space defined by seven trial periods spaced approximately
logarithmically ($P=0.08$, 0.22, 0.87, 1.56, 3.57, 8.30 and 14.72
days), and five planet radii $R_{\rm p}=10.0$, 4.0, 1.0, 0.55, and
0.27\,R$_{\oplus}$.

\begin{figure} 
\plotone{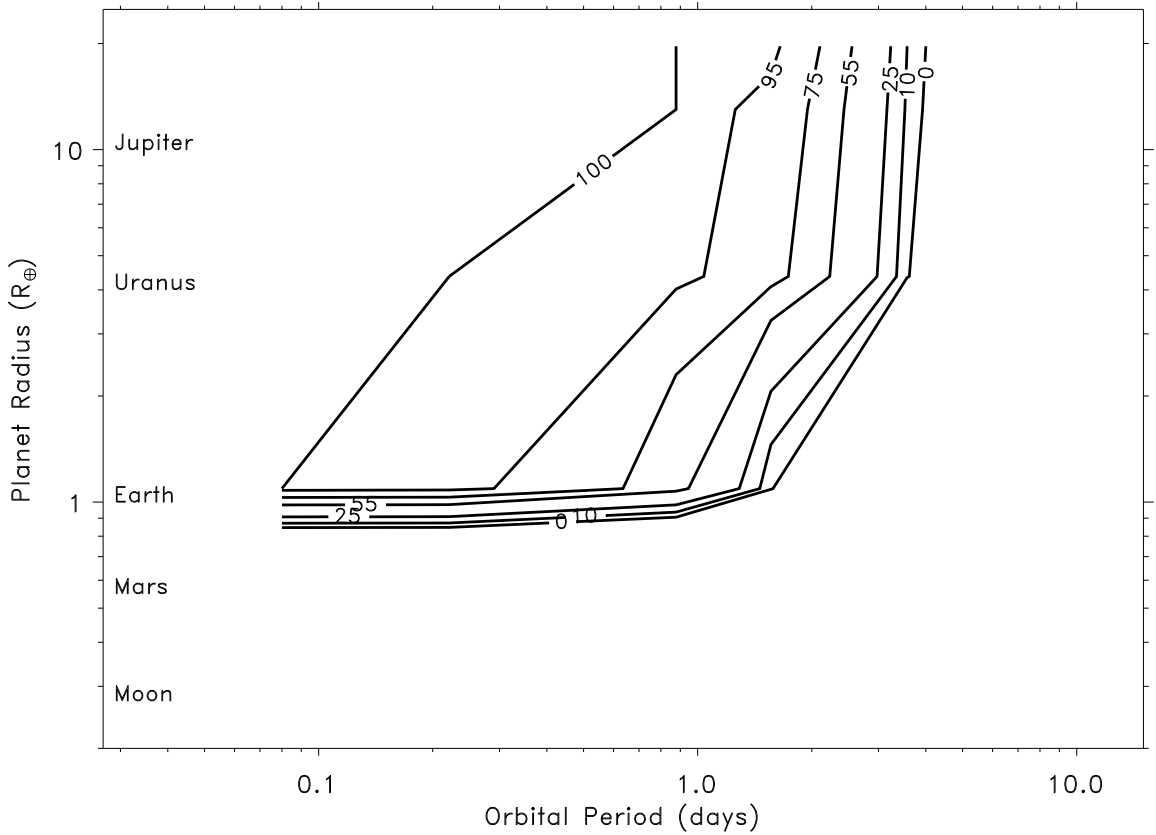}\\
  \caption{\label{fig:f8}Detection limits for planetary companions of sizes ranging
    from Jupiter-size to Moon-size in orbit around the WD1124-293.  The numbers on the contours relate to the detection efficiency in percentage.}
\end{figure}

Figure \ref{fig:f8} shows the detection rate of the simulated transit
signals injected in the WASP light curve of WD1124-293.  We regard as a
match any trial in which the most significant detected period is
within 1\% of being an integer fraction or multiple from $1/5\times$
to $5\times$ the injected transit signal.
From our result on the
simulations conducted using the WASP light curve of WD1124-293 (Figure
\ref{fig:f8}, see also results on simulations conducted on WASP data
detailed in \citealt{Faedi2011}), we would have been able to detect
the transit signal due to a companion of radius 1R$_{\oplus}$ and
period between 2 and 16h with a confidence of 75\%. For objects of
radius similar to that of the Earth but slightly bigger (e.g. up to
3R$_{\oplus}$, or so-called SuperEarths) we would have been able to
detect the eclipse signal for periods between 2h and 12h with a
confidence of 95\% and to about 1d with 75\% confidence. Eclipsing gas
giant planets and substellar objects would have been detectable at
orbital periods between 2h and 1.7d with 95\% confidence, and with
orbital periods of 2d with 75\% confidence. On the other end, no stringent
constraint can be placed on objects with radius $<1 {\rm
  R}_{\oplus}$ even at short periods. The same is true even for large
bodies such as Jupiter with orbital periods longer 2 days. From
Figures 1 and 2 of \citet{agol11} and using the white dwarf
temperature of 9420~K and mass of 0.66 M$_{\odot}$ we find that for
WD1124-293 the white dwarf habitable zone (WDHZ) extends for a region
comprised between 0.02 and 0.045 AU for which, in our simulations, we
can only put weak constraints for Neptune/Uranus-size ($\sim3-4{\rm
  R_{\oplus}}$) and larger objects.  However, eventually WD~1124-293 will cool further and when it cools to below 7000-9000~K, planets with orbital separations of 5$\times10^{-3}$-0.02~AU (P$\sim$0.16-1.3 days) will be in a ``continuous habitable zone'' which lasts for $>$3~Gyr \citep{agol11}.  Our observations thus rule out a significant fraction of $>$ Earth-sized planets in this system that might reside in the continuous habitable zone of WD~1123-293.

We investigated the various factors which affect the
efficiency of detection of these transit signals. When generating each
synthetic light-curve we can easily assess {\it a priori} whether it
will fail the tests requiring a minimum number of individual transits
($>5$) and in-transit data points ($>5$). We find that our
requirements alone render transiting companions essentially
undetectable at our longest trial periods (8.30 and 14.72 days); the
transits are too short in duration and too infrequent to be adequately
sampled. For companions around 1\re\ and larger however there is a
good chance of detection out to periods of $\approx$2 days.

\section{Limits to a variable Accretion Rate to WD~1124-293}
\label{sec:var}
Given the stochastic origins to dust around WDs, it might be possible to observe variability in WD accretion rates.
 Ca variability was claimed to be present in the dusty DAZ G~29-38 \citep{vonhippel07b}, but disputed with other observations \citep{debes08}.  Discrepant abundances of Si between optical and UV observations of some dusty WDs may suggest possible variable accretion \citep{gaensicke12}.  The multiple spectra of WD~1124-293 allow a careful look at the variability in the accretion rate of material onto the surface of the WD through EW measurements of its photospheric Ca~II lines.  WD~1124-293's metal settling timescale for calcium is 194~yr, much longer than the total time high quality spectra of the Ca line have been available, but short enough that small changes due to settling may be observed.  Similarly, if there were any short duration accretion events that briefly increased the accretion rate, we may have observed a sudden increase in the Ca EW. 

Figure \ref{fig:f8} shows the EW as a function of Julian date (JD).  Two measurements of the Ca~II K line, from May 2009 and July 2010, differ significantly from the other measurements.  Similarly, the Ca~II H line measurements from 1998 and   March 2007 also deviate significantly.  The cause of these discrepancies most likely due to lower S/N and poor seeing, but given the good agreement between the other EW measurements and the lack of similar variability at the same time from the complementary Ca~II line, we neglect these measurements in our analysis. To determine whether WD~1124-293's Ca~II line strength was varying, we calculated a $\chi^2$ value for a constant value EW, determined from the median value of the observed EWs.  With this measure, the Ca EW has a $\chi^2$ value of 23 (14) for the Ca~II K (H) line, which corresponds to a probability that the EW is constant to 0.03 (0.3).  Based on our full survey of other DAZs for variability \citep{debmuk10}, we require the probability of constant EW to be less than 10$^{-3}$ to be considered significantly variable.  The standard deviation of our MIKE measurements, neglecting the anomalous measurements, is 4.5\% for the K line and 7\% for the H line.

The roughly constant flow of material over $\sim$11 years for this system places a firm lower limit on the amount of metal-rich material recently deposited, 4.6$\times10^{16}$~g, once again assuming metal rich material at 1\% the solar abundance values, and a total amount of material over one settling time of 9$\times10^{18}$~g assuming a settling time of 194~yr \citep{koester05}.  These masses correspond to asteroids with sizes of 1.5 and 9~km respectively, assuming a bulk density of 3 g/cm$^{3}$.  These radii are comparable to the sizes of asteroids that should survive post-main sequence evolution at an orbital radius of a few AU \citep{jura08}.  

\section{Conclusions}
\label{sec:conc}
We have obtained moderate resolution optical spectroscopy of the DAZ WD~1124-293 over several epochs and found a weak circumstellar absorption feature of Ca~II consistent with a tenuous gaseous disk supplied by small rocky bodies.  Steady state accretion of material onto the WD surface appears to be occurring despite the relatively longer inferred settling timescale for metals, consistent with a recent disruption of an asteroidal body that was relatively small.

WD~1124-293 has also been searched for transiting companions given that any planetary system is nearly edge-on to our line of sight, though none are found for short periods.  Our results do not rule out the possibility for planetary companions at wider separations.  

Emission from gaseous disks have been observed around metal enriched WDs with disks that have been actively accreting dust \citep{gaensicke06,gaensicke07,gaensicke08}.  The origin of that gas could arise from the viscous heating of material \citep{werner09} or from the photoionization of gas \citep{melis10}.  However, these systems may be unusual in the fact that they may represent an early stage in the creation of a dusty disk or represent a massive example of a dusty disk in orbit around a WD.  The accretion of many small planetesimals should proceed with little observable dust, and a primarily gaseous circumstellar disk \citep{jura08}.  These properties match WD~1124-293, representing the first example of a single metal enriched WD with a direct link to the tidal disruption of planetesimals and thus to a relic planetary system.  The large fraction of single WDs that show metal pollution implies a large fraction of post-main sequence objects may house remnant planetary systems with a robust mechanism for delivering small objects in WD-grazing orbits.  A large spectroscopic survey of apparently single, dust-free WDs with high S/N may reveal more systems similar to WD~1124-293.

\acknowledgements
We wish to thank Detlev Koester for unceasing aid in discussing the fundamental parameters of WD~1124-293, and Silvano Desidera for kindly providing valuable spectra of HIP~56280.  We also wish to thank Jean-Rene Roy for discussions on the kinematic broadening of spectral absorption lines, which provided a key impetus for constraining the location of WD~1124-293's gaseous disk.  Finally, we want to thank the anonymous referee, whose attention to detail greatly increased the quality of this paper.  This work made significant use of the SIMBAD and VIZIER databases, operated at CDS, Strasbourg, France.

\bibliography{wd_chap,Faedi}
\bibliographystyle{apj}

\begin{deluxetable}{ccccc}
\tablecolumns{5}
\tablewidth{0pt}
\tablecaption{\label{tab:obs} Spectroscopic Observations}

\tablehead{
\colhead{UT Date (JD-5400000) } & \colhead{Target} &\colhead{Observing Setup} & S/N/pixel (@Ca II) &\colhead{Notes}
}
\startdata
1998-12-12 (51159.629)& WD 1124-293 & Keck/HiRES & 37 & R=30,000 \\
1999-4-19 (51288.328) & WD 1124-293 & Keck/HiRES & 26 & R=30,000 \\
2000-4-23 (51657.663) & WD 1124-293 & VLT/UVES & 36 & R=18500 \\
2000-5-19 (51681.588) & WD 1124-293 & VLT/UVES & 14 & R=18500 \\
2007-3-31 (54189.682) & WD 1124-293 & Clay/MIKE & 24 &  \\
2008-3-21 (54546.605) & WD 1124-293 & Clay/MIKE & 50 & \\
2008-3-22 (54547.656) & WD 1124-293 & Clay/MIKE & 50 & \\
2008-5-13 (54599.536) & WD 1124-293 & Clay/MIKE & 60 & \\
2008-5-14 (54600.551) & WD 1124-293 & Clay/MIKE & 49 & \\
2008-5-15 (54601.477) & WD 1124-293 & Clay/MIKE & 52 & \\
2008-5-16 (54602.502) & WD 1124-293 & Clay/MIKE & 56 & \\
2009-4-16 (54937.686) & WD 1124-293 & Clay/MIKE & 39 & \\
2009-5-18 (54969.548)& WD 1124-293 & Clay/MIKE &  38 &\\
2010-1-1 (55197.727) & HIP 55731 &  Clay/MIKE & 36 & ISM Standard \\
2010-1-1 (55197.733) & HIP 55864 & Clay/MIKE & 40 & ISM Standard \\
2010-1-1 (55197.738) & HIP 55901 &  Clay/MIKE & 100 & ISM Standard \\
2010-1-1 (55197.743) & HIP 55968 &  Clay/MIKE & 47 & ISM Standard \\
2010-6-17 (55365.467) & HIP 55731 &  Clay/MIKE & 78 & ISM Standard \\
2010-6-17 (55365.472) & HIP 55901 & Clay/MIKE &  162 & ISM Standard \\
2010-6-17 (55365.479) & HIP 55864 & Clay/MIKE &  105 & ISM Standard \\
2010-6-17 (55365.486) & WD 1124-293 & Clay/MIKE & 73 & \\
2010-6-17 (55365.575) & HIP 55968 & Clay/MIKE & 56 & ISM Standard \\
2010-7-07 (55384.500) & WD 1124-293 & Clay/MIKE & 33 & \\
2011-6-09 (55385.493) & WD 1124-293 & Clay/MIKE & 70 & \\
\enddata
\end{deluxetable}

\begin{deluxetable}{cccccc}
\tablecolumns{6}
\tablewidth{0pt}
\tablecaption{\label{tab:meas} Properties of Photospheric and Circumstellar Ca~II lines in WD~1124-293}
\tablehead{
\colhead{Line type} &\colhead{JD-5400000} & \colhead{Ca~II K EQW} & \colhead{Ca~II H EQW}  & \colhead{Ca~II K v$_r$} & \colhead{Ca~II H v$_r$} \\
& & \colhead{(m\AA)} & \colhead{(m\AA)} & \colhead{(km s$^{-1}$} & \colhead{(km s$^{-1}$)}
 }
\startdata
 Photospheric & 51159.629 & 104$\pm$ 3 &  39$\pm$ 4 &  30.3 &  31.3 \\
  & 51288.328 &  98$\pm$ 5 &  57$\pm$ 6 &  28.7 &  27.5 \\
 & 51657.663 & 111$\pm$ 6 &  58$\pm$12 &  29.1 &  28.0 \\
 & 51681.588 & 109$\pm$17 &  ... &  29.8 &  ... \\
 & 54189.682 & 106$\pm$ 4 &  34$\pm$ 4 &  29.7 &  34.5 \\
 & 54546.605 & 116$\pm$ 2 &  52$\pm$ 3 &  29.7 &  29.8 \\
 & 54547.658 & 113$\pm$ 2 &  54$\pm$ 3 &  29.0 &  29.4 \\
 & 54599.536 & 113$\pm$ 2 &  53$\pm$ 2 &  28.3 &  28.3 \\
 & 54600.551 & 114$\pm$ 3 &  49$\pm$ 3 &  28.6 &  28.8 \\
 & 54601.477 & 113$\pm$ 2 &  55$\pm$ 2 &  29.0 &  30.1 \\
 & 54602.502 & 118$\pm$ 2 &  55$\pm$ 2 &  29.2 &  29.4 \\
 & 54937.686 & 113$\pm$ 3 &  53$\pm$ 3 &  28.8 &  27.7 \\
 & 54969.548 & 127$\pm$ 3 &  43$\pm$ 3 &  29.1 &  27.8 \\
 & 55364.486 & 115$\pm$ 2 &  52$\pm$ 1 &  28.7 &  29.5 \\
 & 55385.493 &  95$\pm$ 3 &  59$\pm$ 4 &  28.4 &  29.7 \\
 & 55721.598 & 111$\pm$ 2 &  54$\pm$ 2 &  29.7 &  29.4 \\
Circumstellar &  March 2008 & 11$\pm$2 & ... & -0.5$\pm$0.9 & ... \\
 & May 2008 & 12$\pm$1 & ... & -2$\pm$1 & ... \\
  & June 2010+June 2011 & 10$\pm$1 & ... & -1$\pm$1 & ... \\
  & Total & 11$\pm$1 & 2$\pm$0.5 & -0.9$\pm$0.5 & -0.4$\pm$0.5 \\
 \enddata
\end{deluxetable}

\begin{deluxetable}{ccccccccc}
\tablecolumns{9}
\tablewidth{0pt}
\tablecaption{\label{tab:ism}  Properties of ISM clouds for ISM standards}
\tablehead{
\colhead{Name} & \multicolumn{2}{c}{Cloud I} & \multicolumn{2}{c}{Cloud 2} & \multicolumn{2}{c}{Cloud 3}  & \multicolumn{2}{c}{Cloud 4} \\
 & \colhead{v} & \colhead{EW} & \colhead{v} & \colhead{EW} & \colhead{v} & \colhead{EW} & \colhead{v}  & \colhead{EW} \\
 & \colhead{(km s$^{-1}$)} & \colhead{(m\AA)} & \colhead{(km s$^{-1}$)} & \colhead{(m\AA)} & \colhead{(km s$^{-1}$)} & \colhead{(m\AA)} & \colhead{(km s$^{-1}$)} & \colhead{(m\AA)} 
 }
\startdata
HIP~56280 & ... & $<$10 & ... & $<$10 & ... & $<$10 & ... & $<$10 \\
HIP~55864 & ... & $<$ 7  & ... & $<$ 7  & ... & $<$ 7  & ... & $<$ 7  \\
HIP~55731 & ... & $<$ 6  & ... & $<$ 6 & 8.7 & 22$\pm$2 &... & $<$ 6 \\
HIP~55901 & -10.3 & 47$\pm$1 & -1.7 & 34$\pm$1 & 9.5 & 17$\pm$1 & ... & $<$2 \\
HIP~55968 & -6 .3& 72$\pm$2 & 2.0 & 43$\pm$2 & 11.8 & 56$\pm$2 & 21.2 & 19$\pm$2 \\
 \enddata
\end{deluxetable}

\begin{deluxetable}{ccccc}
\tablecolumns{3}
\tablewidth{0pt}
  \tablecaption{Recovery rate of simulated transits of WD1124-293}
  \label{tab:WASP}
\tablehead{  
\colhead{Size}&\colhead{R$_{\rm p}$}&\colhead{$\delta_{\rm tr}$}&\colhead{P}&\colhead{f$_{\rm det}$} \\
      &\colhead{\re}&\colhead{(\%)}&\colhead{(days)}&\colhead{(\%)}\\
      }
\startdata
Jupiter   	&10.0 & 100 &   0.08   & 100  \\
             	&        &        &   0.22   & 100 \\
             	&        &        &   0.87   & 100 \\
             	&        &        &   1.56   &  91  \\
             	&        &        &   3.57   &   8    \\
             	&        &        &   8.30   &   0   \\
\\
Uranus   	& 4.0 & 100 &   0.08   & 100 \\
             	&      &         &   0.22   & 100 \\
             	&      &         &   0.87   &   99 \\
             	&      &         &   1.56   &   82 \\
             	&      &         &   3.57   &   1   \\
             	&      &         &   8.30   &   0   \\
\\							  
Earth	& 1.0 &  49&   0.08  & $-$ \\
          	&       &      &   0.22  &  99 \\
          	&       &      &   0.87  &  61  \\
          	&       &      &   1.56  &   1    \\
          	&       &      &   3.57  &   0   \\
\\							  
				   			   
\enddata
\end{deluxetable}

\begin{figure}
\plotone{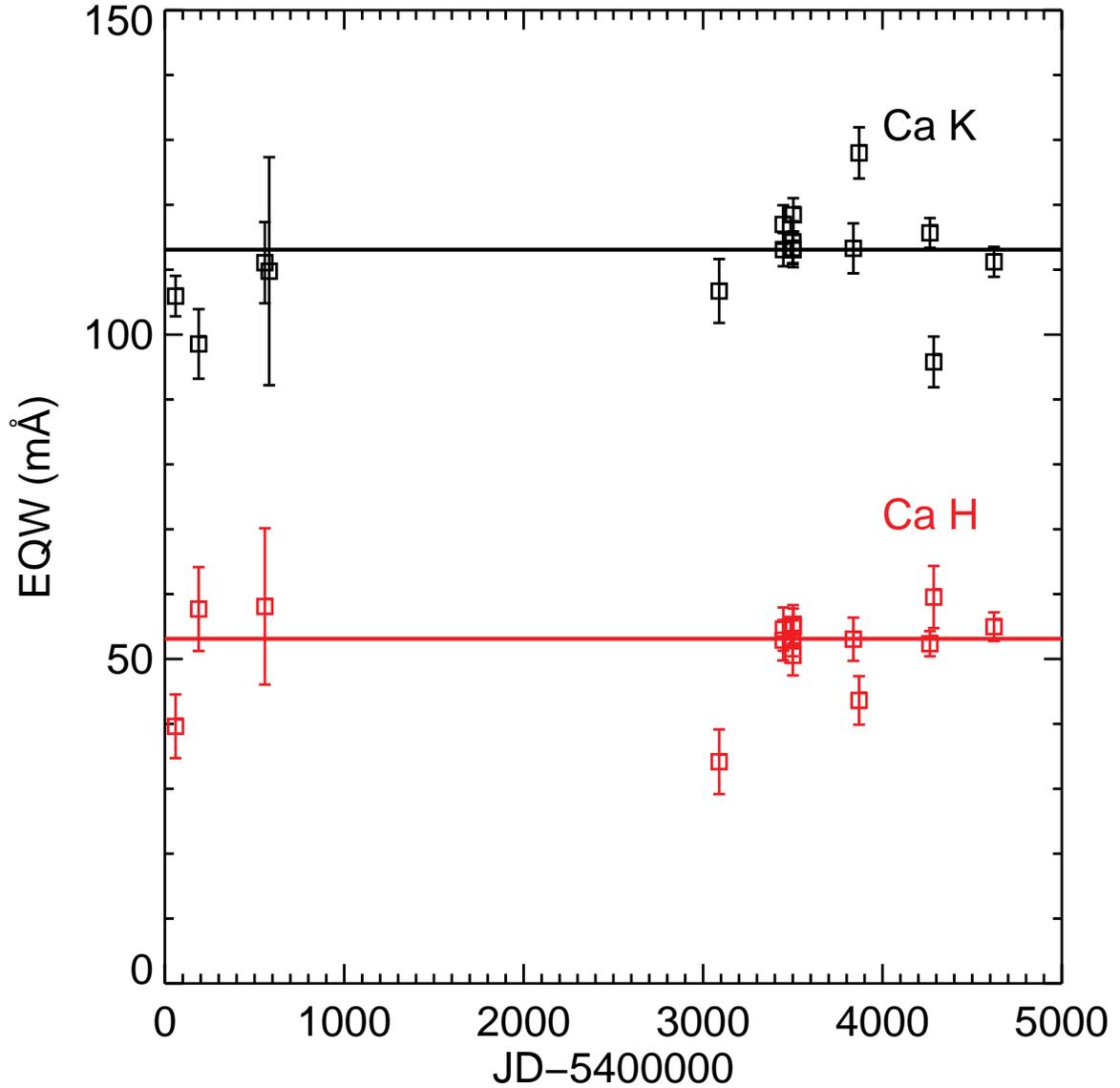}
\caption{\label{fig:f9}The EW of Ca~II in WD~1124-293's photosphere as a function of time.  The EWs are compared to the expectation of a constant value, and do not significantly vary.  Neglecting spurious points, the largest epoch-to-epoch variations are less than 7\% and 5\% from the median EW value for the Ca H (lower points) and K (upper points) lines respectively.}
\end{figure}

\end{document}